\documentclass[a4paper,english]{lipics-v2019}
\usepackage{nth}
\usepackage{microtype}
\usepackage{breqn}
\usepackage{ifthen}
\usepackage{placeins}
\usepackage{complexity}
\usepackage{nicefrac}
\usepackage{todonotes}
\usepackage{siunitx}
\usepackage{amssymb}
\usepackage{color}
\usepackage{gensymb}
\usepackage[noabbrev,capitalise]{cleveref}
\usepackage{mathtools}

\hideLIPIcs
\nolinenumbers

\newcommand{\old}[1]{{}}

\title{Maximum Polygon Packing:\\ The CG:SHOP Challenge 2024}
\titlerunning{Maximum Polygon Packing: CG Challenge 2024}

\author{Sándor P.~Fekete}{Department of Computer Science, TU Braunschweig, Germany}{s.fekete@tu-bs.de}{https://orcid.org/0000-0002-9062-4241}{}
\author{Phillip Keldenich}{Department of Computer Science, TU Braunschweig, Germany}{p.keldenich@tu-bs.de}{https://orcid.org/0000-0002-6677-5090}{}
\author{Dominik Krupke}{Department of Computer Science, TU Braunschweig, Germany}{d.krupke@tu-bs.de}{https://orcid.org/0000-0003-1573-3496}{}
\author{Stefan Schirra}{Department for Simulation and Graphics, OvGU Magdeburg, Germany}{stschirr@isg.cs.uni-magdeburg.de}{https://orcid.org/0009-0006-5928-1494}{}
\authorrunning{S.~P.~Fekete, P.~Keldenich, D.~Krupke, S.~Schirra}
\Copyright{S.~P.~Fekete, P.~Keldenich, D.~Krupke, S.~Schirra}



\supplement{}

\funding{Work at TU Braunschweig has been partially supported by the German Research Foundation (DFG), 
project ``Computational Geometry: Solving Hard Optimization Problems'' (CG:SHOP), grant FE407/21-1.} 

\acknowledgements{We thank the members of the CG:SHOP Challenge Advisory Board for their valuable input: 
William J.~Cook, Andreas Fabri, Dan Halperin, Michael Kerber, Philipp Kindermann, Joe Mitchell, Kevin Verbeek.}

\EventAcronym{CG:SHOP 2024}

\begin{document}
\maketitle
\begin{abstract}
We give an overview of the 2024 Computational Geometry Challenge
targeting the problem \textsc{Maximum Polygon Packing}:
Given a convex region $P$ in the plane, and a collection of simple polygons $Q_1, \ldots, Q_n$, each $Q_i$
with a respective value $c_i$,
find a subset $S \subseteq \{1, \ldots,n\}$
and a feasible packing within $P$ of the polygons $Q_i$ (without rotation) for $i \in S$,
maximizing $\sum_{i \in S} c_i$.
Geometric packing problems, such as this, present significant computational challenges and are of substantial practical importance.
\end{abstract}

\section{Introduction}
The ``CG:SHOP Challenge'' (Computational Geometry: Solving Hard
Optimization Problems) originated as a workshop at the 2019
Computational Geometry Week (CG Week) in Portland, Oregon in June,
2019.  The goal was to conduct a computational challenge competition
that focused attention on a specific hard geometric optimization
problem, encouraging researchers to devise and implement solution
methods that could be compared scientifically based on how well they
performed on a database of carefully selected and varied instances.
While much of computational
geometry research is theoretical, often seeking provable approximation
algorithms for \NP-hard optimization problems,
the goal of the Challenge was to set the metric of success based on
computational results on a specific set of benchmark geometric
instances. The 2019 Challenge~\cite{CGChallenge2019_JEA} focused on the problem of computing
simple polygons of minimum and maximum area for given sets of vertices in the
plane. It generated a strong response from many research
groups~\cite{area-crombez,area-tau,area-salzburg,area-exact,area-campinas,area-omega} from both the computational geometry and the combinatorial
optimization communities, and resulted in a lively exchange of
solution ideas.

Subsequently, the CG:SHOP Challenge became an event within the CG Week
program, with top performing solutions reported in the Symposium on
Computational Geometry (SoCG) proceedings. 
The schedule for the Challenge was
advanced earlier, to give an opportunity for more participation, particularly
among students, e.g., as part of course projects. 
For CG Weeks 2020, 2021, 2022, and 2023,  the Challenge problems were \textsc{Minimum Convex Partition}~\cite{CGChallenge2020,SoCG2020_1,SoCG2020_2,SoCG2020_3}, 
\textsc{Coordinated Motion Planning}~\cite{CGChallenge2021,Challenge2021_1,Challenge2021_2,Challenge2021_3,SoCG2021_1,SoCG2021_2,SoCG2021_3},
\textsc{Minimum Partition into Plane Subgraphs}~\cite{CGChallenge2022_JEA,Challenge2022_1,Challenge2022_2,SoCG2022_1,SoCG2022_2,SoCG2022_3,SoCG2022_4},
and \textsc{Minimum Convex Covering}~\cite{fekete2023minimum,Challenge2023_1,Challenge2023_2},
respectively.  

The sixth edition of the Challenge in 2024 continued
this format, leading to contributions in the SoCG proceedings.

\section{The Challenge: Maximum Polygon Packing}

A suitable contest problem has a number of desirable properties.

\begin{itemize}
\item The problem is of geometric nature.
\item The problem is of general scientific interest and has received previous attention.
\item Optimization problems tend to be more suitable than feasibility problems; in principle, 
  feasibility problems are also possible, but they need to be suitable for sufficiently
  fine-grained scoring to produce an interesting contest.
\item Computing optimal solutions is difficult for instances of reasonable size.
\item This difficulty is of a fundamental algorithmic nature, and not only due to
 issues of encoding or access to sophisticated software or hardware.
\item Verifying feasibility of provided solutions is relatively easy.
\end{itemize}

In this sixth year, a call for suitable problems was communicated in March
2023. In response, a number of interesting problems were proposed for the 2024
Challenge. These were evaluated with respect to difficulty, distinctiveness
from previous years, and existing literature and related work. In the end, the
Advisory Board selected the chosen problem. Special thanks go to Mikkel Abrahamsen
(University of Copenhagen) who suggested this problem, motivated by a rich history
in geometry and optimization, including~\cite{LTWYC1990packing} and a wide range of
previous work described further down.

\subsection{The Problem}
The specific problem that formed the basis of the 2024 CG Challenge was the following. 

\medskip
\noindent
\textbf{Problem:} \textsc{Maximum Polygon Packing}\\
\textbf{Given:} A convex region $P$ in the plane, and a collection of simple polygons $Q_1, \ldots, Q_n$, each $Q_i$ 
with a respective value $c_i$. \\
\textbf{Goal:} Find a subset $S \subseteq \{1, \ldots,n\}$
and a feasible packing within $P$ of the polygons $Q_i$ (without rotation) for $i \in S$,
maximizing $\sum_{i \in S} c_i$.

\subsection{Related Work}
Problems of geometric packing have been studied for a long time, with classic
challenges such as the Kepler conjecture~\cite{Kepler,kepler_05,kepler_17} on
densest sphere packings.  Providing a survey that does justice to the wide
range of relevant work goes beyond the scope of this overview;
we refer to Fejes Tóth~\cite{toth1999recent, toth20172}, Lodi, Martello and Monaci~\cite{LMM2002two},
Brass, Moser and Pach~\cite{brass2006research} and Böröczky~\cite{boroczky2004finite} for more comprehensive surveys
on the theoretical side, and Dyckhoff and Finke~\cite{dyckhoff1992cutting},
Sweeney and Paternoster~\cite{sweeney1992cutting},
Bennell et al.~\cite{bennell2009tutorial,bennell2008geometry},
Leao et al.~\cite{leao2020irregular} on the practical side.

Even the decision problem whether it is possible to pack a given set of (not necessarily identical) axis-aligned 
squares into the unit
square was shown to be strongly $\NP$-complete by Leung et
al.~\cite{LTWYC1990packing}, using a reduction from \textsc{3-Partition}.
Additional difficulties arise when
rotation of the squares is allowed, even when packing unit squares into 
a square~\cite{erdos1975packing,gensane2005improved,chung2020efficient}.
For packing polygons, 
Allen and Iacono~\cite{allen2012packing} showed that even packing a maximum
number of \emph{identical} simple polygons into a square is 
NP-complete. For packing a set of nonidentical polygons,
Abrahamsen et al.~\cite{abrahamsen2020framework} gave a general
framework for establishing $\exists\mathbb{R}$-hardness, making it unlikely
that the problem belongs to NP. 

Despite these theoretical difficulties, there is also a wide range of positive
results. Already in 1967, Moon and Moser~\cite{MM1967some} proved that it is possible to
pack a set of squares into the unit square if their total area does not exceed
$\nicefrac{1}{2}$, which is best possible. 
More recently, Merino and Wiese~\cite{merino2020two} considered two-dimensional
Knapsack Problems for weighted convex polygons, which is closely related
to our Challenge problem. They gave a number of approximation algorithms,
even in the presence of rotation. From the context of practical Computational 
Geometry, Milenkovic et
al.~\cite{daniels1997multiple,milenkovic1999translational,milenkovic1997multiple,milenkovic1999rotational,milenkovic1998rotational,milenkovic1996translational}
provided a spectrum of heuristic methods, while Fekete and
Schepers~\cite{schepers_exact,fs-gfbhd-04,schepers_esa,fs-hdpm-97,fs-hdpb-97,fs-hdpea-97} developed exact
methods for optimally packing axis-aligned orthogonal objects in two and higher dimensions.




\subsection{Instances}

Creating suitable instances is a critical aspect of any challenge.
Instances that are too easy to solve undermine the challenge's complexity, making it trivial.
Conversely, instances that demand extensive computational resources for basic
preprocessing or for finding viable solutions can provide an unfair advantage to teams with
superior computing facilities.  This issue is exacerbated when the instance set
is too voluminous to be managed by a small number of computers.

We utilized four distinct instance generators for the contest, each permitting the adjustment of various parameters, such as item complexity or variance.
This approach ensures a diverse collection of instances. The generators are described as follows.

\begin{figure}
	\centering
	\includegraphics[width=0.7\textwidth]{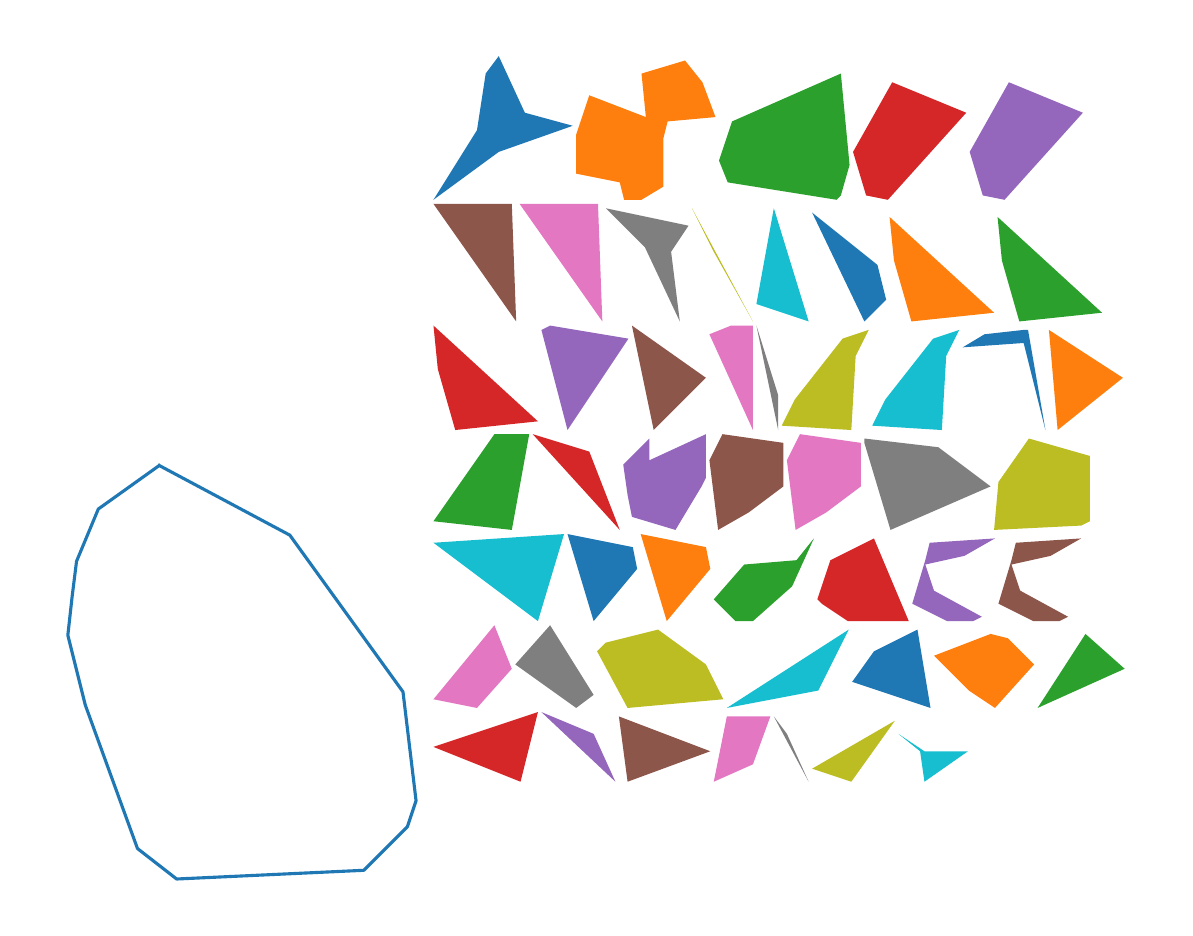}
	\caption{An example of a \textbf{random} instance.}
	\label{fig:random-instance}
\end{figure}

\begin{figure}
	\centering
	\includegraphics[width=0.7\textwidth]{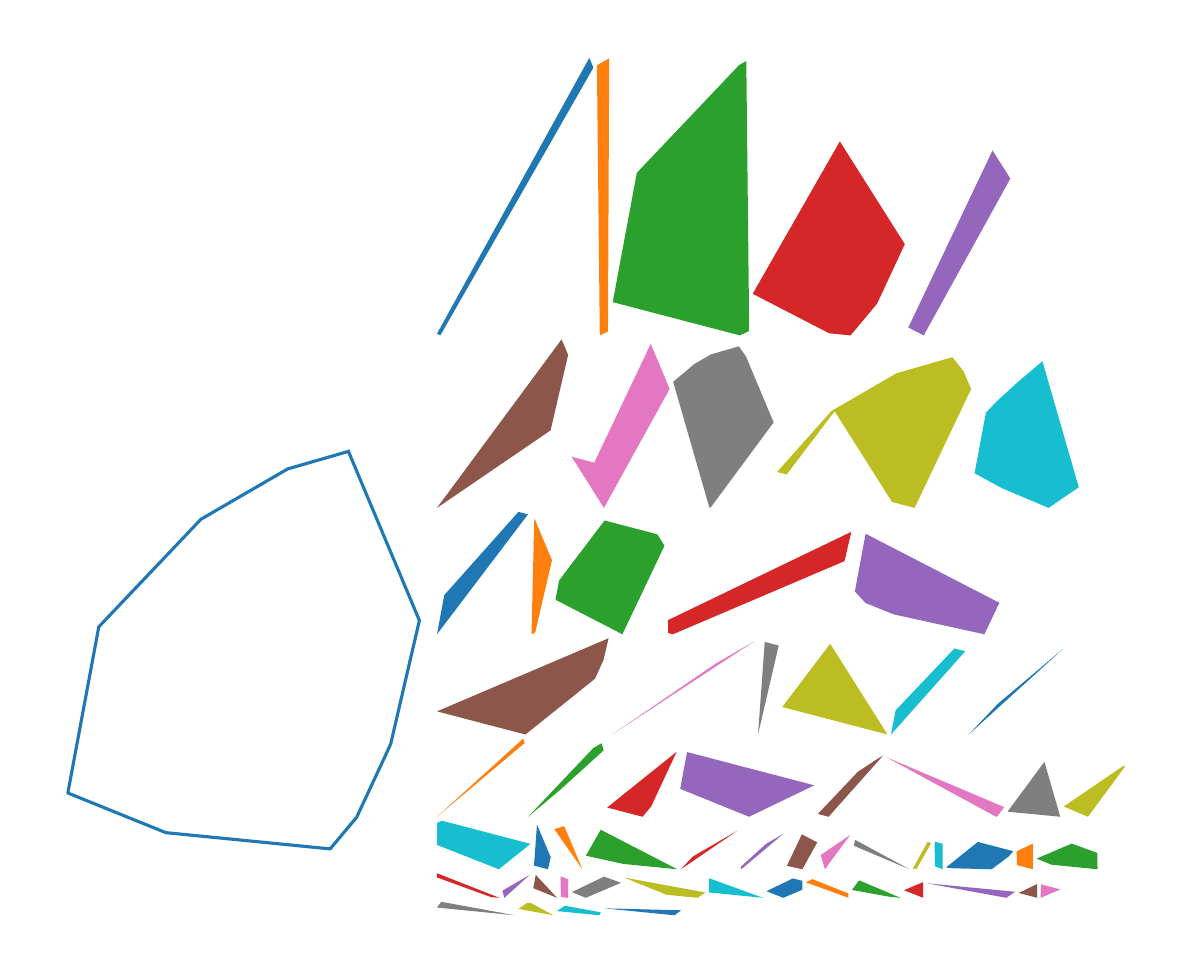}
	\caption{An example of a \textbf{jigsaw} instance.}
	\label{fig:jigsaw-instance}
\end{figure}

\begin{figure}
	\centering
	\includegraphics[width=0.7\textwidth]{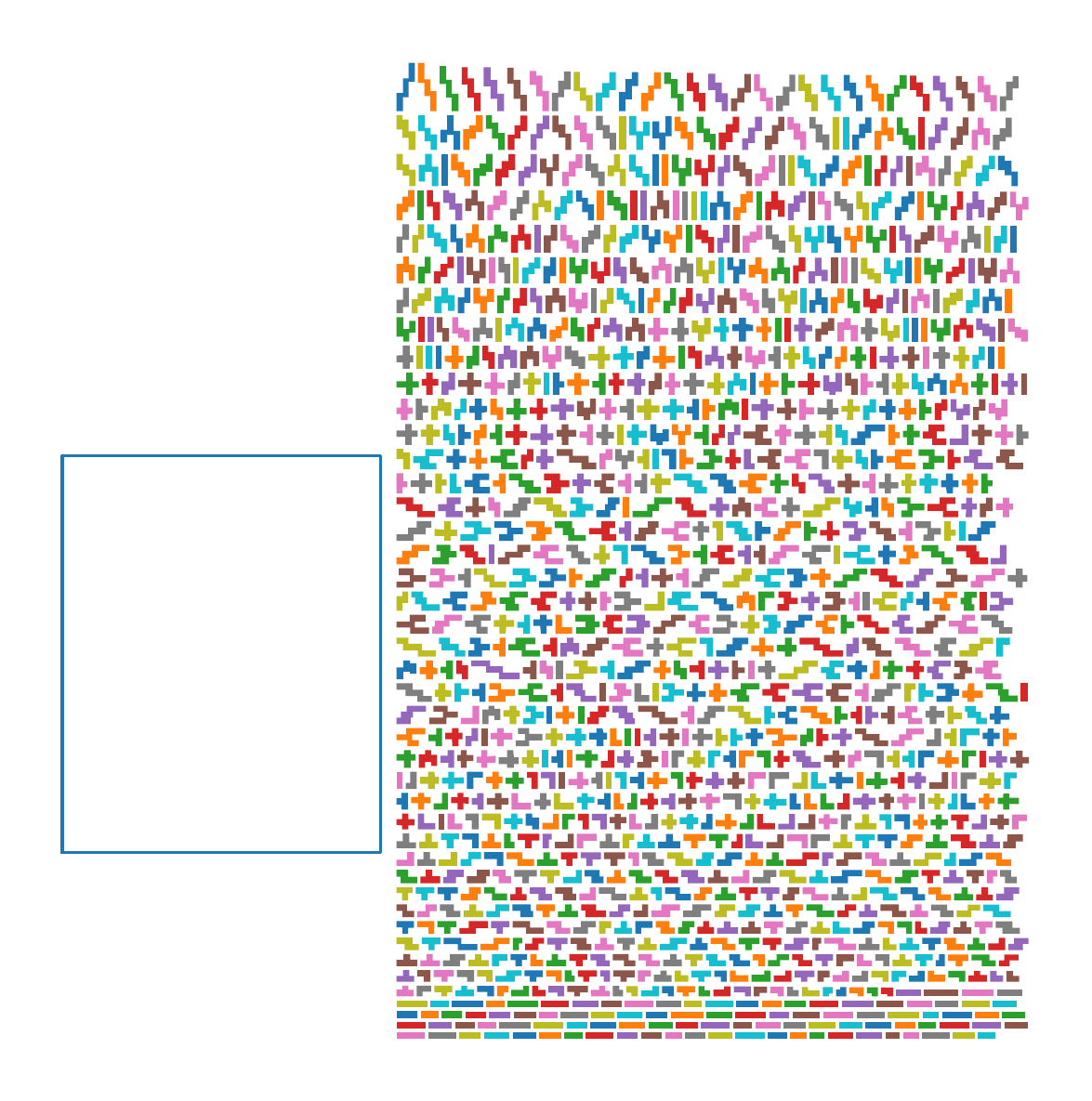}
	\caption{An example of an \textbf{atris} instance.}
	\label{fig:atris-instance}
\end{figure}

\begin{figure}
	\centering
	\includegraphics[width=0.65\textwidth]{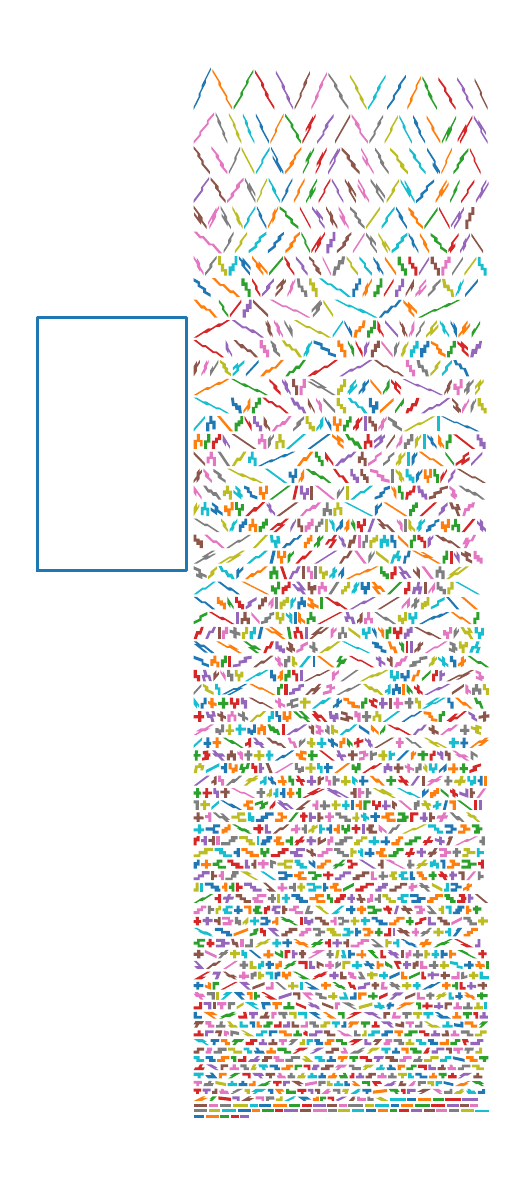}
	\caption{An example of a \textbf{satris} instance.}
	\label{fig:satris-instance}
\end{figure}

\begin{description}
	\item[random] Instances are created by generating a set of random polygons and a random container. 
		The process involves generating several random points and then forming either a convex or a concave hull (determined by a random ratio) to form polygons,
		enabling the production of instances with items whose sizes and complexities adhere to configurable distributions.
		See \cref{fig:random-instance} for an example.
    \item[jigsaw] These instances originate from segmenting the container or its bounding box into pieces, then slightly altering the pieces to hinder easy reassembly.
	The segmentation is performed by first adding a number of random lines through the container or its bounding box.
	This results only in convex and often relatively simple pieces, such that in a second step, we go through the geometric arrangement and randomly combine adjacent faces to form more complex pieces.
	During this process, we make sure not to create non-simple polygons or generally badly shaped or sized pieces.
	Given the nature of the Knapsack Problem, an item surplus is generated by amalgamating multiple jigsaws, thereby forming a complex instance. See \cref{fig:jigsaw-instance} for an example.
  \item[atris] The \textbf{atris} generator uses a rectangular strip of some specified width and height as the container.
               It generates items that are derived from polyominos in the well-known computer game Tetris.
               More specifically, the generator creates items from seven different categories: \emph{Line}, \emph{Squiggly}, \emph{Double Squiggly},
               \emph{Y}, \emph{T}, \emph{L} and \emph{Plus}.

               However, unlike in Tetris, the width and height of the {pixels} constituting these items are not fixed, but individual columns and rows of pixels
               have their width and height chosen uniformly at random between some minimum and maximum value.
               Furthermore, for items consisting of multiple arms such as a \emph{Y}, a \emph{Plus} or a \emph{Squiggly},
               each of the arms can be shifted by some randomly chosen distance, such that the resulting shape need not be symmetric. 
               This shifting is restricted to prevent disconnecting the item or change it to another type of shape.

               Each item can be flipped vertically and horizontally based on a fair coin toss;
               furthermore, each item is rotated by an integer multiple of $90\degree$, chosen between $0$ and $3$ uniformly at random.

               Item generation stops once the total item area exceeds some preselected multiple $t \in [1,2]$ of the container area;
               instances with a larger number of items use larger containers.

               The value of each item is derived by multiplying a random scaling factor from $[0.8,1.2]$ with the 
               item's area, further multiplying the result with a constant depending on the item's shape category, giving more weight to items that are 
               arguably harder to pack (such as Y-shapes and Double Squigglies) and less weight to easier shapes such as rectangles.
               The resulting value is then rounded to an integer.
               
               We took care to avoid any possible overflow or imprecision when working with the scores; all scores and coordinates generated by this generator
               are integral and even for the largest contest instances, the sum of all item scores was below $2^{40}$,
               well below the threshold at which \texttt{double} values are no longer able to precisely encode integers.

			   See \cref{fig:atris-instance} for an example.
              
  \item[satris] The \textbf{satris} generator works nearly identically to the \textbf{atris} generator; in addition,
                items are randomly selected for an additional shearing transformation, i.e., they have their coordinates
                transformed according to a matrix \(\begin{psmallmatrix}1 & m\\ 0 & 1\\ \end{psmallmatrix},\) with $m \in [0.1,2]$ chosen uniformly at random;
                this transformation happens before the flipping and rotating transformation outlined above.

                Sheared items have all their coordinates rounded to the nearest integer. 
                Their value is increased compared to non-sheared items, both by multiplication with a fixed constant 
                and by multiplication with another constant increasing with $m$, so items that are more severely sheared receive higher value.

				See \cref{fig:satris-instance} for an example.
\end{description}


In addition to the geometric shapes of the polygons, the value attributed to each polygon plays a crucial role in the problem instances.
To construct our instances, we employed several value functions, outlined as follows.

\begin{description}
    \item[Area:] The value of a polygon is directly proportional to its area.
    \item[Convex Hull Area:] A polygon's value corresponds to the area of its convex hull.
    \item[Rotated Bounding Box:] The value is based on the area of the polygon's smallest enclosing rectangle.
	\item[Uniform:] Each polygon is assigned a uniform value, regardless of its size.
\end{description}

To introduce further variability into our instances, we incorporated noise into the value functions for certain cases.
To this end, the value is adjusted by multiplying it with a minor random factor, encouraging a strategic approach to item prioritization beyond mere density optimization.
The specific value functions utilized were not revealed to the challenge participants.

\begin{figure}
	\centering
	\includegraphics[width=0.65\textwidth]{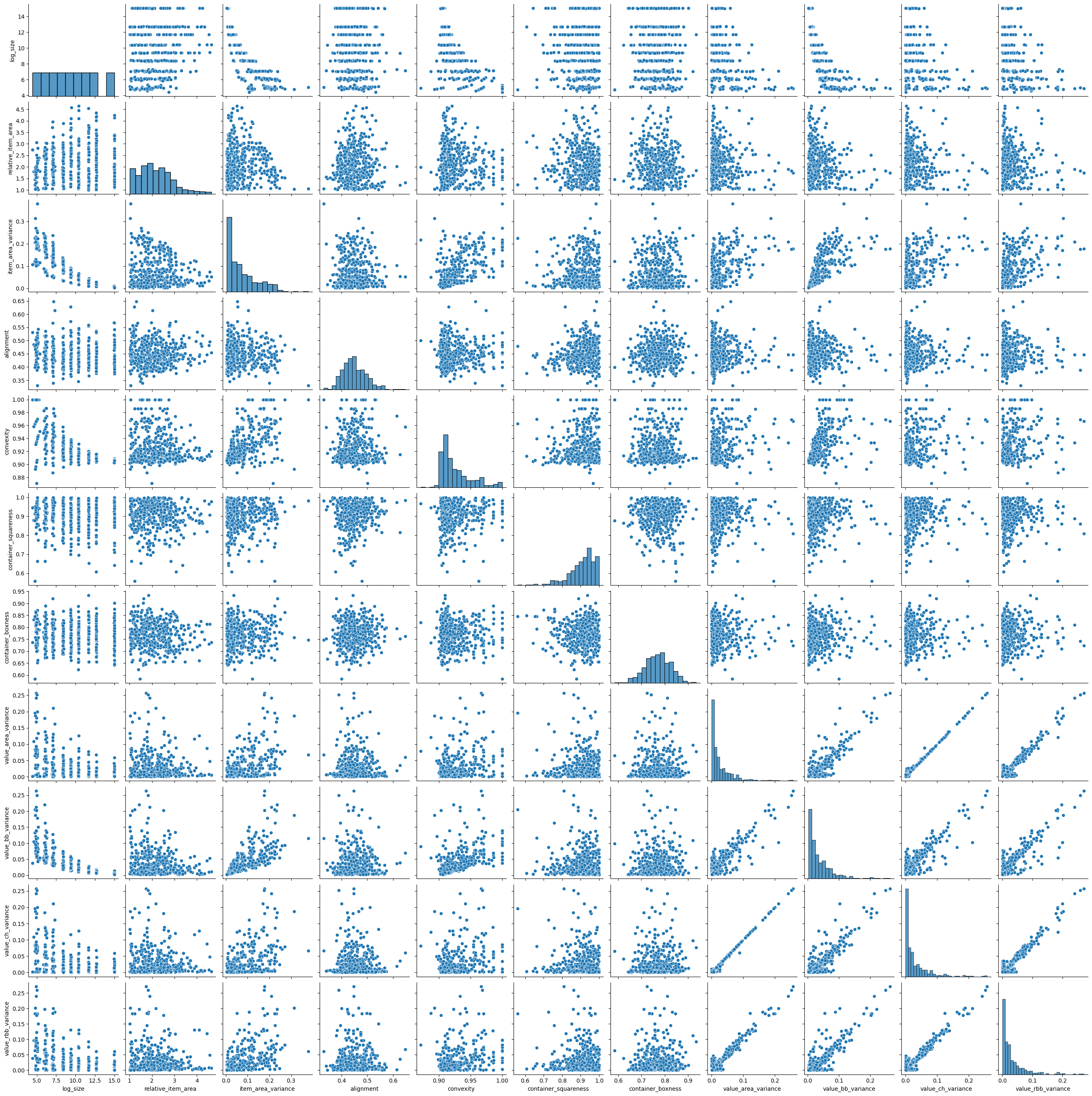}
	\\[1cm]
	\includegraphics[width=0.65\textwidth]{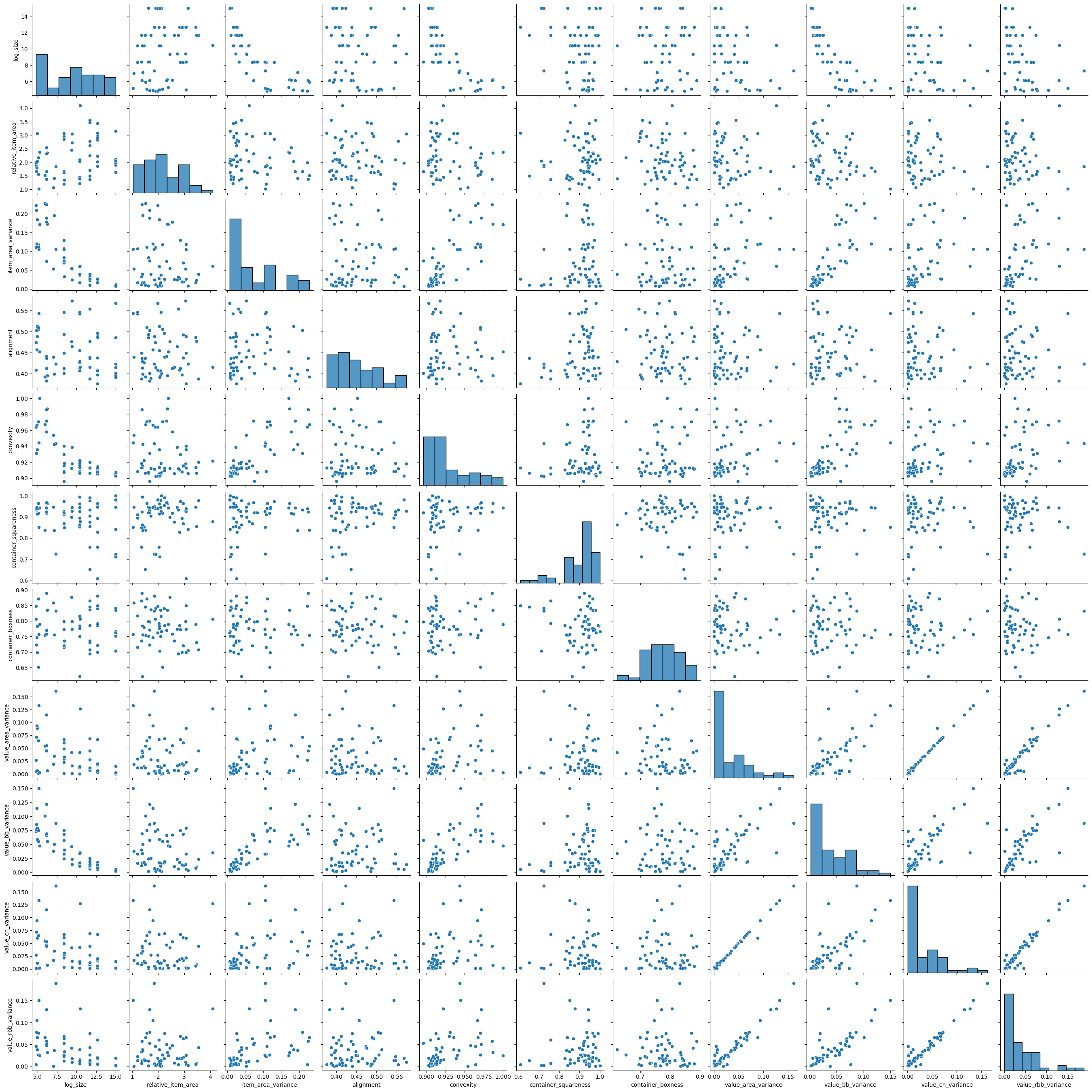}
	\caption{Pair plots of the instance metrics for the initial set of instances (top) and the final benchmark (bottom), showing the correlation between the different metrics and the distribution of the instances among them.}
	\label{fig:instance-selection}
\end{figure}

Using the described generators and value functions, we created a large set of instances, of which a subset was selected for the contest.
This subset was selected by first identifying a set of eleven metrics that describe the instances and allow quantifying their differences.
These metrics included the logarithm of the number of items, the difference of the area to the area of the convex hull, the difference to a (rotated) rectangle of container and items, the relation of area of items to the container area, a rough measure on how aligned the items are, and further.
As these metrics have some correlations and just computing the geometric differences would put 
too much weight on specific aspects even after normalization,
we used a principal component analysis (PCA) to reduce the dimensionality, and thus correlation, of the instance space.
To select a diverse set of instances of a specific size, we then grouped similar instances into the desired number of groups using a k-means clustering algorithm, and selected a random instance from each group for the final benchmark.
The instance distribution of the initial and final set of instances can be seen in \cref{fig:instance-selection}.
A more detailed description of this approach can be found in Chapter~10 of \cite{Dissertation:Krupke}.

\subsection{Evaluation}

The competition featured a total of 180 instances.
For each instance \(I\), define \(B(I)\) as the highest value achieved by any team's best solution for \(I\).
Similarly, let \(T(I)\) represent the value of team \(T\)'s best solution for \(I\).
The score \(S_T(I)\) for team \(T\) on instance \(I\) is calculated as:
\[ S_T(I) := \frac{{T(I)}^2}{{B(I)}^2}. \]

Thus, if a team's value is half that of the best solution, their score for this instance reduces to $0.25$.
Teams achieving the highest submitted value for \(I\) receive a perfect score of $1$ for that instance.
Teams failing to submit any valid solution for \(I\) are assigned a score of 0, corresponding to a solution that fails to pack any items into the container.
The overall score \(S_T\) for each team \(T\) is the sum of \(S_T(I)\) across all instances.
The contest winner is the team with the highest total score.
In the event of a tie, the deciding factor would have been the time at which the tied score was first achieved; 
just like in the past, this tiebreaker was not needed.

The benefit of employing a relative scoring system lies in its independence
from the quality of bounds or approximations, instead directly comparing team
performances.  While scoring relative to a lower or upper
bound offers consistency, the impact of an instance on the overall score may
significantly depend on the bound's accuracy.
This approach could inadvertently diminish the influence of challenging and
engaging instances---often characterized by less precise bounds---on the
total score.

\subsection{Categories}

The contest was run in an \emph{Open Class}, in which participants could use any
computing device, any amount of computing time (within the duration of the
contest) and any team composition. 

\subsection{Server and Timeline}

The competition was facilitated through a dedicated server at TU Braunschweig,
 accessible at \url{https://cgshop.ibr.cs.tu-bs.de/competition/cg-shop-2024/}.
An initial batch of example instances was made available on July 31, 2023,
followed by the release of the final benchmark set on September 29, 2023. 
The competition concluded on January 22, 2024 (AoE).

Participants were provided with a verification tool as an open-source Python package, available at \url{https://github.com/CG-SHOP/pyutils24}.
This tool provided detailed information on errors and allowed participants to easily investigate and correct their solutions if the server rejected them.
To ensure accurate verification and address potential floating point arithmetic issues, the tool leveraged the CGAL library~\cite{cgal}.
Given that many instances feature items of similar sizes, leading to a homogeneous distribution in optimal packings, the adoption of a simple quad tree structure significantly accelerated the verification process.
This approach enabled the selective examination of nearby items for intersections, facilitating nearly instantaneous solution verification, even in the presence of the computational demands of exact arithmetic.

\section{Outcomes}

\Cref{table:competition_scores} outlines the outcomes for the participating teams, ranked by their performance.
Overall, nine teams participated and agreed to be evaluated and ranked. Four teams reached scores above \num{150} of a possible \num{180} and were invited for contributions in the 2024 SoCG proceedings, as follows.

\begin{enumerate}
\item Team Shadoks: Guilherme Dias da Fonseca and Yan Gerard, ``Shadoks Approach to Knapsack Polygonal Packing''~\cite{Challenge24_1}.
\item Team SmartPlacer: Canhui Luo, Zhouxing Su and Zhipeng Lü, ``A General Heuristic Approach for Maximum Polygon Packing''~\cite{Challenge24_2}.
\item Team CGA Lab Salzburg: Martin Held, ``Priority-Driven Nesting of Irregular Polygonal Shapes Within a Convex Polygonal Container Based on a Hierarchical Integer Grid''~\cite{Challenge24_3}.
\item Team TU Dortmund: Alkan Atak, Kevin Buchin, Mart Hagedoorn, Jona Heinrichs, Karsten Hogreve, Guangping Li, Patrick Pawelczyk, ```Computing Maximum Polygonal Packings in Convex Polygons using Best-Fit, Genetic Algorithms and ILPs''~\cite{Challenge24_4}.
\end{enumerate}

\begin{table}[t]
	\centering
	\begin{tabular}{|c|l|r|r|r|r|}
	\hline
	\textbf{Rank} & \textbf{Team}            & \textbf{Score} & \# best & \# only $\geq 0.95$ & \# only $\geq 0.9$           \\ \hline
	1             & Shadoks                  & 173.91        & 71 & 32 &12           \\
	2             & SmartPlacer              & 167.67        & 58 & 28 &  2           \\
	3             & CGA Lab Salzburg         & 156.62        & 9 & 3 & 0           \\
	4             & TU Dortmund              & 154.91        & 37 & 19 & 8           \\
	5             & AMW                      & 146.64        & 1 & 0 & 0           \\
	6             & GreedyPackers            & 128.63        & 0 & 0 & 0           \\
	7             & polygonizers             & 89.06         & 0 & 0 & 0             \\
	8             & QTeam                    & 72.59         & 0 & 0 & 0             \\
	9             & CodingAtChristmas        & 3.39          & 0 & 0 & 0             \\ \hline
	\end{tabular}
	\caption{Competition scores of the teams that submitted at least one valid solution and opted to be evaluated and ranked.
		The columns \# best, \# only $\geq 0.95$, and \# only $\geq 0.9$ show the number of instances for which the team achieved the best score, as only team a score of at least $0.95$, and as only team a score of at least $0.9$, respectively.
	}
	\label{table:competition_scores}
\end{table}

Team Shadoks used Integer Linear Programming (ILP) and a carefully designed greedy approach for 
initial solutions, which were then improved by local search. Smart Placers 
subdivided large problems into smaller subproblems, for which they tried to fill 
a subcontainer sequentially, using a combination of local and tabu 
search to arrive at conflict-free placements. CGA Lab Salzburg used a 
hierarchical grid and priority heuristics for filling the container 
polygon sequentially. TU Dortmund's approach involved different 
strategies depending on problem size: ILPs were used for small, genetic 
algorithms for medium and a best-fit approach for large instances.

\begin{figure}
	\centering
	\includegraphics[width=\textwidth]{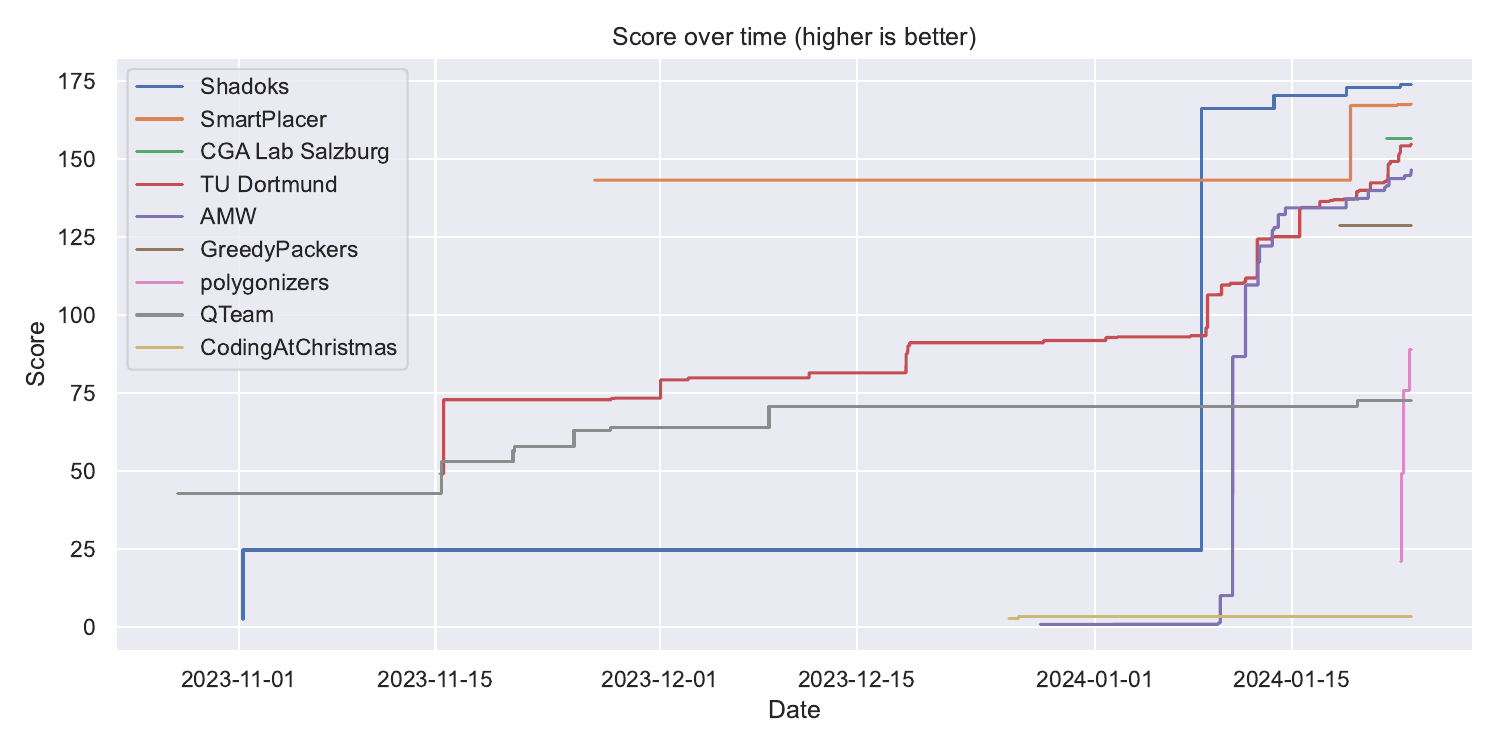}
	\caption{
		The plot traces the evolution of solution quality throughout the contest.
		The overall score, derived from the sum of scores for individual instances based on final submissions,
		illustrates a steady progression. Notably, team TU Dortmund exhibits incremental improvements,
		culminating in a score that more than doubles. CGA Lab Salzburg makes a late yet impactful entry.
		The final two weeks witness significant activity across teams.
		}
		\label{fig:score_over_time}
\end{figure}

The top-ranking team, Shadoks, achieved an outstanding score of 173.91, leading the competition.
Close behind, SmartPlacer scored 167.67.
CGA Lab Salzburg and TU Dortmund follow with scores of 156.62 and 154.91, respectively.
AMW and GreedyPackers also demonstrated commendable efforts with scores just above 146 and 128, respectively.
The progress over time of each team’s score can be seen in Figure~\ref{fig:score_over_time}; overviews of best solutions are shown
in the subsequent figures. 

\begin{figure}
	\centering
	\includegraphics[width=\textwidth]{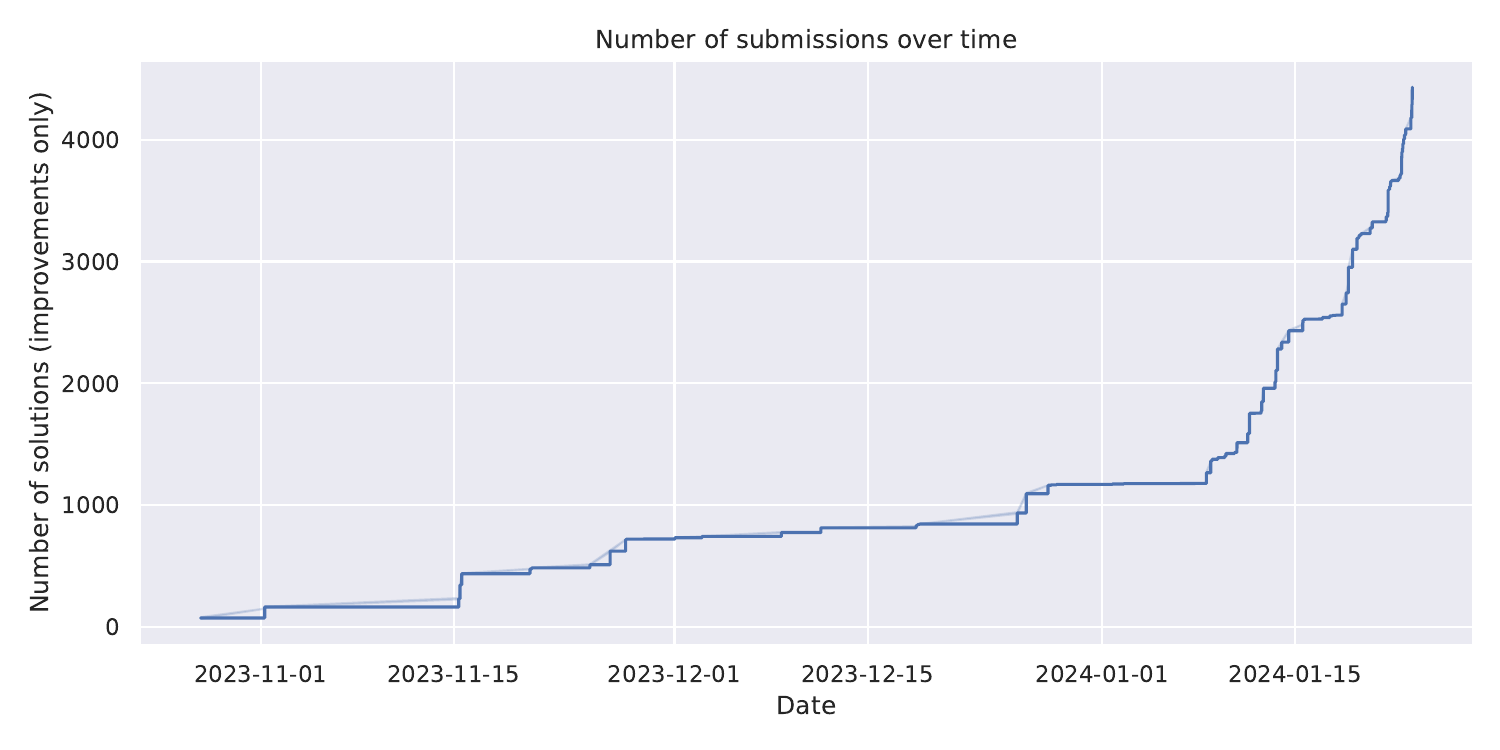}
	\caption{
		This plot depicts the number of submissions over time,
		highlighting a concentration of activity in the final two weeks.
		Unlike in previous Challenge iterations, the surge in submissions on the final day is less pronounced.
		}
		\label{fig:submissions_over_time}
\end{figure}

\begin{figure}
	\centering
	\includegraphics[width=\textwidth]{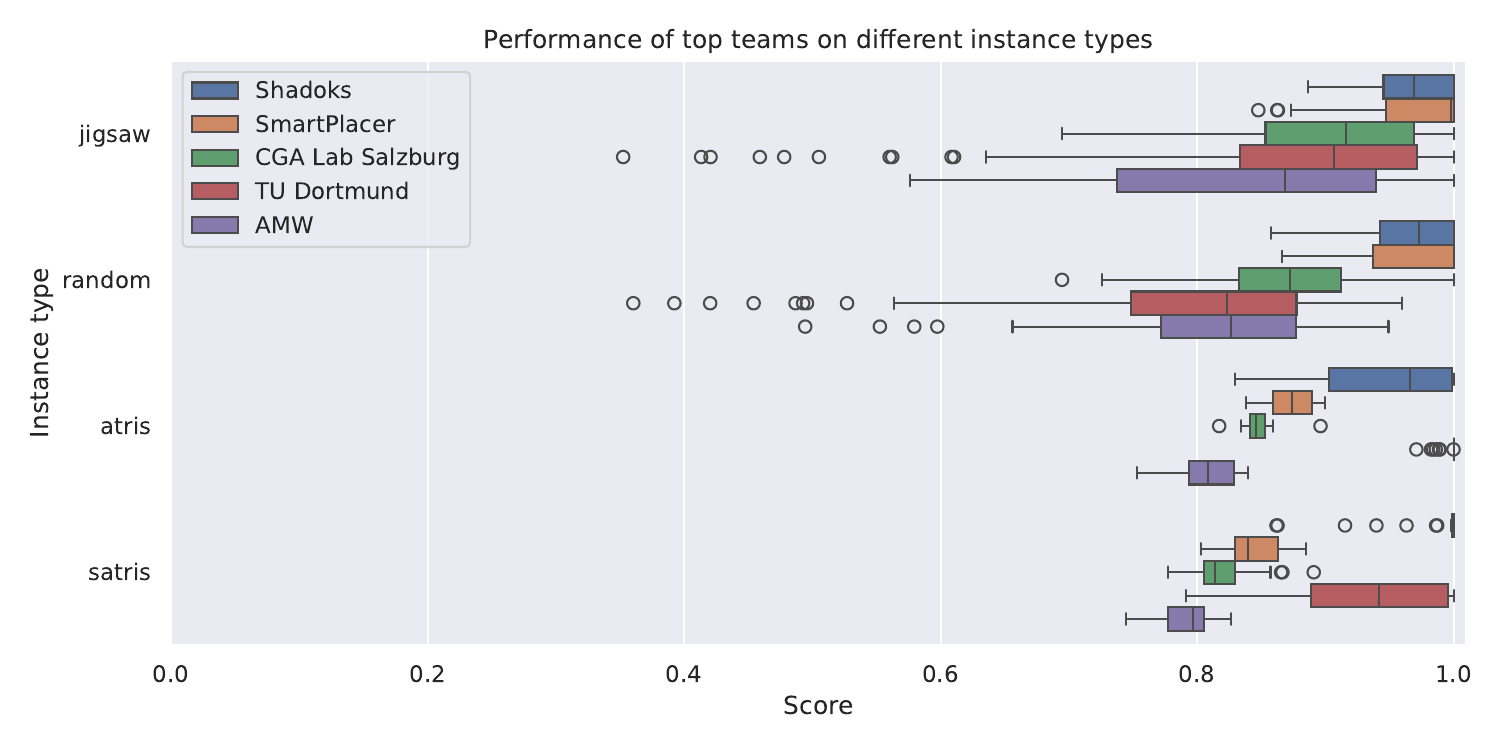}
	\caption{
		This plot contrasts the performance of the top five teams across different instance types, illustrating the distribution of scores.
		As the scores are relative to the best solution, this allows for a straightforward comparison of team performances.
		The box plot depicts the median, quartiles, and outliers. Here, an outlier is specified as a score lying beyond 1.5 times the interquartile range above the upper quartile or below the lower quartile, offering insights into the variability and central tendency of each team's scores.
		We can see that team Shadoks leads in \textbf{satris} instances, while TU Dortmund excels in \textbf{atris} instances.
		Team SmartPlacers shows strength in \textbf{jigsaw} and \textbf{random} instances,
		albeit with weaker performance elsewhere.
		CGA Lab Salzburg and AMW exhibit no particular preference for instance types.
	}
	\label{fig:performance_of_top_teams_on_different_instance_types}
\end{figure}

\begin{figure}
	\centering
	\includegraphics[width=\textwidth]{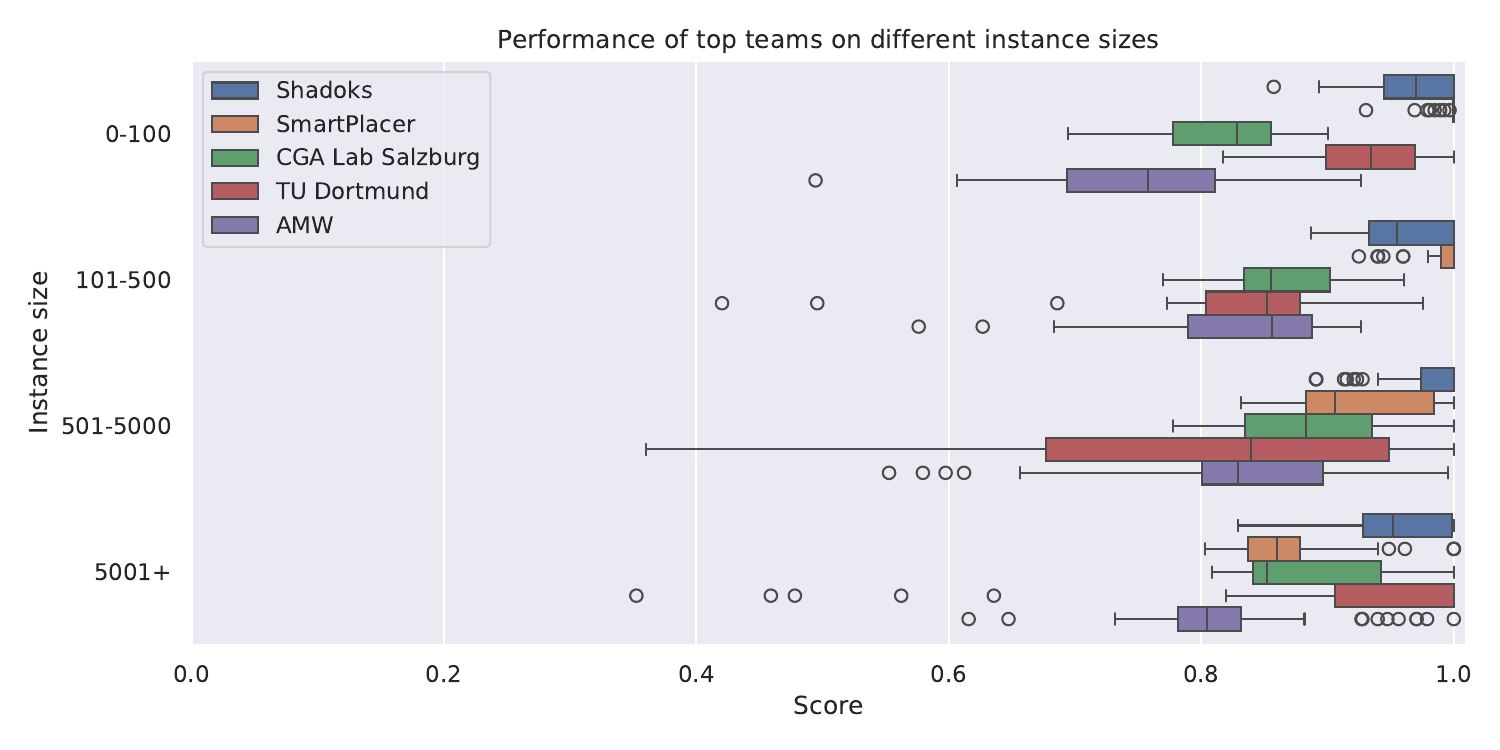}
	\caption{
		This plot examines how the top five teams fare across different instance sizes.
		SmartPlacer dominates instances with up to 500 items.
		The team Shadoks emerge as the front-runner for instances sized between 501 and 5000 items.
		No distinctive patterns of dominance are observed for CGA Salzburg and AMW.
		Interestingly, TU Dortmund shows a dip in performance for medium-sized instances, but rebounds strongly for larger instances.
	}
	\label{fig:top_teams_performance_instance_size}
\end{figure}

\begin{figure}
	\centering
	\includegraphics[width=\textwidth]{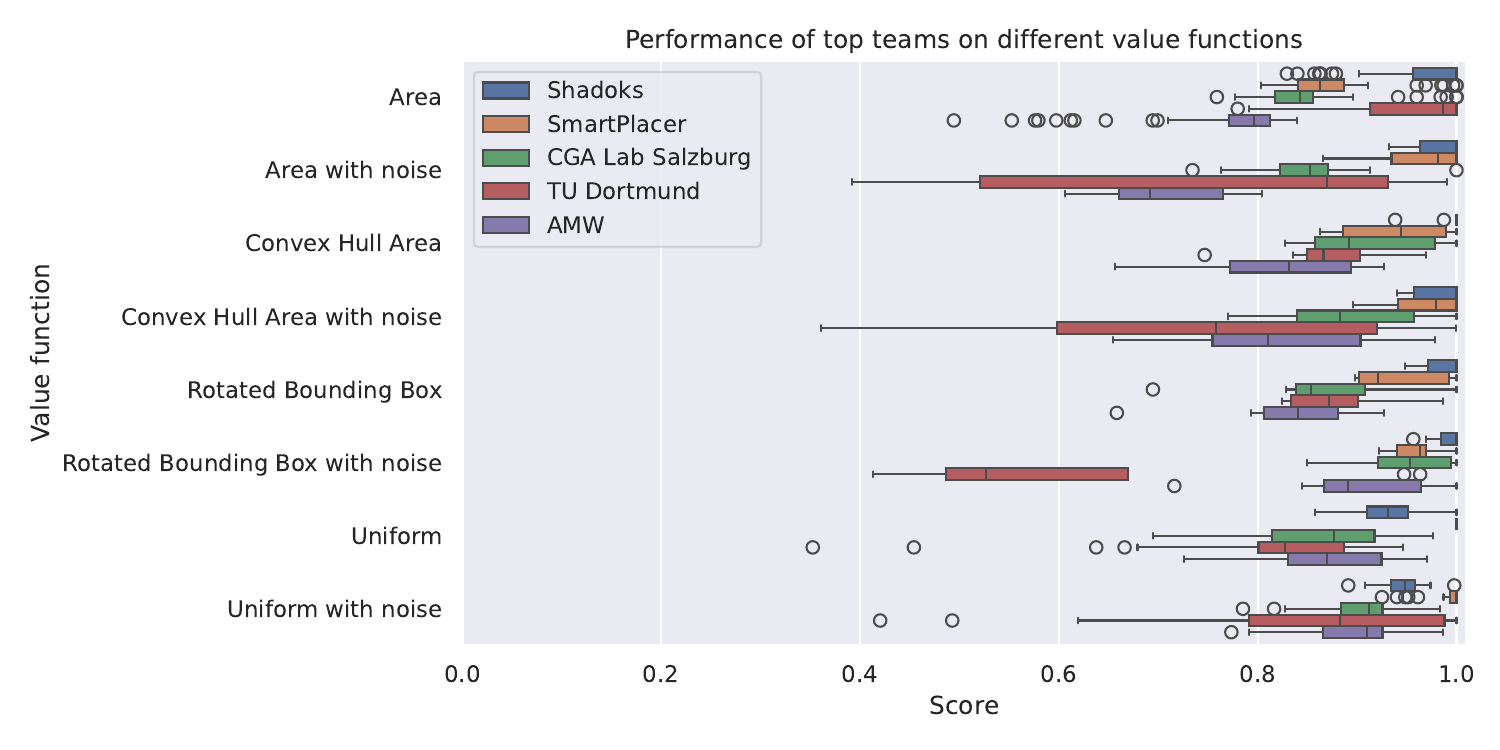}
	\caption{
		This plot assesses the impact of different value functions on the scores of the top five teams.
		TU Dortmund experiences a decline in performance with noise-infused value functions.
		Team Shadoks slightly falters in scenarios for which items have uniform values, a domain for which SmartPlacer excels.
		Shadoks, however, dominates in instances for which value functions are determined by the area of the convex hull.
	}
	\label{fig:top_teams_performance_instance_value_type}
\end{figure}

The competition data reveals insightful trends in team performance across various metrics, as shown in the accompanying figures.
\Cref{fig:score_over_time} illustrates the dynamic evolution of solution quality, highlighting continuous improvements by team TU Dortmund and a strategic late entry by CGA Lab Salzburg, with a surge in activity in the final weeks.
\Cref{fig:submissions_over_time} indicates a significant increase in submissions as the contest neared its end, albeit with a less pronounced last-day spike compared to previous competitions.
Performance analysis across different instance types in \cref{fig:performance_of_top_teams_on_different_instance_types} shows team Shadoks excelling in \textbf{satris} instances, TU Dortmund in \textbf{atris}, and SmartPlacers demonstrating strength in \textbf{jigsaw} and 
\textbf{random} instances, with CGA Lab Salzburg and AMW maintaining balanced performances.
\Cref{fig:top_teams_performance_instance_size} analyzes performance trends relative to instance size, with SmartPlacer dominating 
in smaller instances, Shadoks leading in larger ones, and TU Dortmund displaying a notable rebound in performance for the largest instances.
Finally, \cref{fig:top_teams_performance_instance_value_type} assesses the impact of value functions on team scores, highlighting the challenges faced by TU Dortmund with noise-infused values, the proficiency of SmartPlacer with uniform values, and the dominance of Shadoks in area-based value functions.
Collectively, these figures highlight the complexities and strategic nuances of the competition, reflecting the diverse 
strengths and adaptabilities of the participating teams, and hint at the perspectives for further improvements.

Note that this analysis may be influenced by correlations due to the uneven distribution of metrics.
For instance, atris and satris instances exclusively use area as their value function,
whereas random and jigsaw instances incorporate all four value functions.
This discrepancy could affect the interpretation of results and comparisons across different instance types.
You can investigate the solutions of the top teams for individual instances on \url{https://pageperso.lis-lab.fr/guilherme.fonseca/cgshop24view/} (developed by the team Shadoks).

\section{Conclusions}
The 2024 CG:SHOP Challenge motivated a considerable number of teams to engage
in extensive optimization studies. The outcomes promise further insight into
the underlying, important optimization problem. This demonstrates the
importance of considering geometric optimization problems from a practical 
perspective.

\bibliography{bibliography,references}

\begin{thebibliography}{10}

\bibitem{Challenge2023_2}
M.~Abrahamsen, W.~B. Meyling, and A.~Nusser.
\newblock Constructing concise convex covers via clique covers.
\newblock In {\em Symposium on Computational Geometry (SoCG)}, volume 258 of
  {\em LIPIcs}, pages 66:1--66:9, 2023.

\bibitem{abrahamsen2020framework}
M.~Abrahamsen, T.~Miltzow, and N.~Seiferth.
\newblock Framework for er-completeness of two-dimensional packing problems.
\newblock In {\em Symposium on Foundations of Computer Science (FOCS)}, pages
  1014--1021, 2020.

\bibitem{allen2012packing}
S.~R. Allen and J.~Iacono.
\newblock Packing identical simple polygons is {NP}-hard.
\newblock {\em arXiv preprint arXiv:1209.5307}, 2012.

\bibitem{Challenge24_4}
A.~Atak, K.~Buchin, M.~Hagedoorn, J.~Heinrichs, K.~Hogreve, G.~Li, and
  P.~Pawelczyk.
\newblock Computing maximum polygonal packings in convex polygons using
  best-fit, genetic algorithms and ilps.
\newblock In {\em Symposium on Computational Geometry (SoCG)}, volume 293 of
  {\em LIPIcs}, pages 86:1--86:9, 2024.

\bibitem{bennell2008geometry}
J.~A. Bennell and J.~F. Oliveira.
\newblock The geometry of nesting problems: A tutorial.
\newblock {\em European Journal of Operational Research}, 184(2):397--415,
  2008.

\bibitem{bennell2009tutorial}
J.~A. Bennell and J.~F. Oliveira.
\newblock A tutorial in irregular shape packing problems.
\newblock {\em Journal of the Operational Research Society}, 60:S93--S105,
  2009.

\bibitem{boroczky2004finite}
K.~B{\"o}r{\"o}czky~Jr.
\newblock {\em Finite {P}acking and {C}overing ({C}ambridge {T}racts in
  {M}athematics)}.
\newblock Cambridge University Press, 2004.

\bibitem{brass2006research}
P.~Brass, W.~O.~J. Moser, and J.~Pach.
\newblock {\em Research Problems in Discrete Geometry}.
\newblock Springer Science \& Business Media, 2005.

\bibitem{chung2020efficient}
F.~Chung and R.~Graham.
\newblock Efficient packings of unit squares in a large square.
\newblock {\em Discrete \& Computational Geometry}, 64(3):690--699, 2020.

\bibitem{Challenge2022_1}
L.~Crombez, G.~D. da~Fonseca, F.~Fontan, Y.~Gerard, A.~Gonzalez{-}Lorenzo,
  P.~Lafourcade, L.~Libralesso, B.~Mom{\`{e}}ge, J.~Spalding{-}Jamieson,
  B.~Zhang, and D.~W. Zheng.
\newblock Conflict optimization for binary {CSP} applied to minimum partition
  into plane subgraphs and graph coloring.
\newblock {\em Journal of Experimental Algorithms}, 28:1.2:1--1.2:13, 2023.

\bibitem{area-crombez}
L.~Crombez, G.~D. da~Fonseca, and Y.~Gerard.
\newblock Greedy and local search solutions to the minimum and maximum area.
\newblock {\em Journal of Experimental Algorithmics}, 27:2.2:1--2.2:11, 2022.

\bibitem{SoCG2022_1}
L.~Crombez, G.~D. da~Fonseca, Y.~Gerard, and A.~Gonzalez{-}Lorenzo.
\newblock Shadoks approach to minimum partition into plane subgraphs.
\newblock In {\em Symposium on Computational Geometry (SoCG)}, volume 224 of
  {\em LIPIcs}, pages 71:1--71:8, 2022.

\bibitem{SoCG2021_1}
L.~Crombez, G.~D. da~Fonseca, Y.~Gerard, A.~Gonzalez{-}Lorenzo, P.~Lafourcade,
  and L.~Libralesso.
\newblock Shadoks approach to low-makespan coordinated motion planning.
\newblock In {\em Symposium on Computational Geometry (SoCG)}, volume 189 of
  {\em LIPIcs}, pages 63:1--63:9, 2021.

\bibitem{Challenge2021_1}
L.~Crombez, G.~D. da~Fonseca, Y.~Gerard, A.~Gonzalez{-}Lorenzo, P.~Lafourcade,
  and L.~Libralesso.
\newblock Shadoks approach to low-makespan coordinated motion planning.
\newblock {\em Journal of Experimental Algorithms}, 27:3.2:1--3.2:17, 2022.

\bibitem{Challenge2021_2}
L.~Crombez, G.~D. da~Fonseca, Y.~Gerard, A.~Gonzalez{-}Lorenzo, P.~Lafourcade,
  and L.~Libralesso.
\newblock Shadoks approach to low-makespan coordinated motion planning.
\newblock {\em Journal of Experimental Algorithms}, 27:3.2:1--3.2:17, 2022.

\bibitem{Challenge2023_1}
G.~D. da~Fonseca.
\newblock Shadoks approach to convex covering.
\newblock In {\em Symposium on Computational Geometry (SoCG)}, volume 258 of
  {\em LIPIcs}, pages 67:1--67:9, 2023.

\bibitem{Challenge24_1}
G.~D. da~Fonseca and Y.~Gerard.
\newblock Shadoks approach to knapsack polygonal packing.
\newblock In {\em Symposium on Computational Geometry (SoCG)}, volume 293 of
  {\em LIPIcs}, pages 83:1--83:9, 2024.

\bibitem{daniels1997multiple}
K.~Daniels and V.~J. Milenkovic.
\newblock Multiple translational containment, {Part I}: An approximate
  algorithm.
\newblock {\em Algorithmica}, 19(1):148--182, 1997.

\bibitem{CGChallenge2020}
E.~D. Demaine, S.~P. Fekete, P.~Keldenich, D.~Krupke, and J.~S.~B. Mitchell.
\newblock Computing convex partitions for point sets in the plane: The {CG:SHOP
  Challenge} 2020, 2020.

\bibitem{dyckhoff1992cutting}
H.~Dyckhoff and U.~Finke.
\newblock {\em Cutting and packing in production and distribution: A typology
  and bibliography}.
\newblock Springer Science \& Business Media, 1992.

\bibitem{SoCG2020_3}
G.~Eder, M.~Held, S.~de~Lorenzo, and P.~Palfrader.
\newblock Computing low-cost convex partitions for planar point sets based on
  tailored decompositions.
\newblock In {\em Symposium on Computational Geometry (SoCG)}, volume 164 of
  {\em LIPIcs}, pages 85:1--85:11, 2020.

\bibitem{area-salzburg}
G.~Eder, M.~Held, S.~Jasonarson, P.~Mayer, and P.~Palfrader.
\newblock 2-opt moves and flips for area-optimal polygonalizations.
\newblock {\em Journal of Experimental Algorithmics}, 27:2.7:1--2.7:12, 2022.

\bibitem{erdos1975packing}
P.~Erd{\"o}s and R.~L. Graham.
\newblock On packing squares with equal squares.
\newblock {\em Journal of Combinatorial Theory, Series A}, 19(1):119--123,
  1975.

\bibitem{toth1999recent}
G.~Fejes~T\'{o}th.
\newblock Recent progress on packing and covering.
\newblock {\em Contemporary Mathematics}, 223:145--162, 1999.

\bibitem{area-exact}
S.~P. Fekete, A.~Haas, P.~Keldenich, M.~Perk, and A.~Schmidt.
\newblock Computing area-optimal simple polygonalization.
\newblock {\em Journal of Experimental Algorithmics}, 27:2.6:1--2.6:23, 2020.

\bibitem{CGChallenge2019_JEA}
S.~P. Fekete, P.~Keldenich, D.~Krupke, and J.~S.~B. Mitchell.
\newblock Area-optimal simple polygonalizations: The {CG Challenge} 2019.
\newblock {\em Journal of Experimental Algorithms}, 27:1--12, 2022.

\bibitem{CGChallenge2021}
S.~P. Fekete, P.~Keldenich, D.~Krupke, and J.~S.~B. Mitchell.
\newblock Computing coordinated motion plans for robot swarms: The {CG:SHOP
  Challenge} 2021.
\newblock {\em Journal of Experimental Algorithms}, 27:3.1:1--3.1:12, 2022.

\bibitem{CGChallenge2022_JEA}
S.~P. Fekete, P.~Keldenich, D.~Krupke, and S.~Schirra.
\newblock Minimum partition into plane subgraphs: The {CG:}{SHOP} challenge
  2022.
\newblock {\em Journal of Experimental Algorithms}, 28:1.9:1--1.9:13, 2022.

\bibitem{fekete2023minimum}
S.~P. Fekete, P.~Keldenich, D.~Krupke, and S.~Schirra.
\newblock Minimum coverage by convex polygons: The {CG:SHOP Challenge} 2023,
  2023.

\bibitem{schepers_esa}
S.~P. Fekete and J.~Schepers.
\newblock A new exact algorithm for general orthogonal d-dimensional knapsack
  problems.
\newblock In {\em European Symposium on Algorithms (ESA)}, pages 144--156,
  1997.

\bibitem{fs-hdpm-97}
S.~P. Fekete and J.~Schepers.
\newblock On higher-dimensional packing {I}: Modeling.
\newblock Technical Report 97--288, Universit{\"a}t zu K{\"o}ln, 1997.

\bibitem{fs-hdpb-97}
S.~P. Fekete and J.~Schepers.
\newblock On higher-dimensional packing {II}: Bounds.
\newblock Technical Report 97--289, Universit{\"a}t zu K{\"o}ln, 1997.

\bibitem{fs-hdpea-97}
S.~P. Fekete and J.~Schepers.
\newblock On higher-dimensional packing {III}: Exact algorithms.
\newblock Technical Report 97--290, Universit{\"a}t zu K{\"o}ln, 1997.

\bibitem{fs-gfbhd-04}
S.~P. Fekete and J.~Schepers.
\newblock A general framework for bounds for higher-dimensional orthogonal
  packing problems.
\newblock {\em Mathematical Methods of Operations Research}, 60:311--329, 2004.

\bibitem{schepers_exact}
S.~P. Fekete, J.~Schepers, and J.~van~der Veen.
\newblock An exact algorithm for higher-dimensional orthogonal packing.
\newblock {\em Operations Research}, 55(3):569--587, 2007.

\bibitem{SoCG2022_3}
F.~Fontan, P.~Lafourcade, L.~Libralesso, and B.~Mom{\`{e}}ge.
\newblock Local search with weighting schemes for the {CG:} {SHOP} 2022
  competition.
\newblock In {\em Symposium on Computational Geometry (SoCG)}, volume 224 of
  {\em LIPIcs}, pages 73:1--73:6, 2022.

\bibitem{gensane2005improved}
T.~Gensane and P.~Ryckelynck.
\newblock Improved dense packings of congruent squares in a square.
\newblock {\em Discrete \& Computational Geometry}, 34(1):97--109, 2005.

\bibitem{area-tau}
N.~Goren, E.~Fogel, and D.~Halperin.
\newblock Area-optimal polygonization using simulated annealing.
\newblock {\em Journal of Experimental Algorithmics}, 27:2.3:1--2.3:17, 2022.

\bibitem{kepler_17}
T.~Hales, M.~Adams, G.~Bauer, T.~D. Dang, J.~Harrison, L.~T. Hoang,
  C.~Kaliszyk, V.~Magron, S.~McLaughlin, T.~T. Nguyen, Q.~T. Nguyen, T.~Nipkow,
  S.~Obua, J.~Pleso, J.~Rute, A.~Solovyev, T.~H.~A. Ta, N.~T. Tran, T.~D.
  Trieu, J.~Urban, K.~Vu, and R.~Zumkeller.
\newblock A formal proof of the {K}epler conjecture.
\newblock {\em Forum of Mathematics, Pi}, 5:e2, 2017.

\bibitem{kepler_05}
T.~C. Hales.
\newblock A proof of the {K}epler conjecture.
\newblock {\em Annals of Mathematics}, 162(3):1065--1185, 2005.

\bibitem{Challenge24_3}
M.~Held.
\newblock Priority-driven nesting of irregular polygonal shapes within a convex
  polygonal container based on a hierarchical integer grid.
\newblock In {\em Symposium on Computational Geometry (SoCG)}, volume 293 of
  {\em LIPIcs}, pages 85:1--85:6, 2024.

\bibitem{Kepler}
J.~Kepler.
\newblock Strena seu de nive sexangula, 1611.

\bibitem{Dissertation:Krupke}
D.~M. Krupke.
\newblock {\em Algorithm Engineering for Hard Problems in Computational
  Geometry}.
\newblock PhD thesis, May 2022.

\bibitem{leao2020irregular}
A.~A. Leao, F.~M. Toledo, J.~F. Oliveira, M.~A. Carravilla, and
  R.~Alvarez-Vald{\'e}s.
\newblock Irregular packing problems: A review of mathematical models.
\newblock {\em European Journal of Operational Research}, 282(3):803--822,
  2020.

\bibitem{area-omega}
J.~Lepagnot, L.~Moalic, and D.~Schmitt.
\newblock Optimal area polygonization by triangulation and ray-tracing.
\newblock {\em Journal of Experimental Algorithmics}, 27:1--23, 2022.

\bibitem{LTWYC1990packing}
J.~Y.~T. Leung, T.~W. Tam, C.~S. Wong, G.~H. Young, and F.~Y.~L. Chin.
\newblock Packing squares into a square.
\newblock {\em Journal of Parallel and Distributed Computing}, 10(3):271–275,
  1990.

\bibitem{SoCG2021_2}
P.~Liu, J.~Spalding{-}Jamieson, B.~Zhang, and D.~W. Zheng.
\newblock Coordinated motion planning through randomized k-opt.
\newblock In {\em Symposium on Computational Geometry (SoCG)}, volume 189 of
  {\em LIPIcs}, pages 64:1--64:8, 2021.

\bibitem{LMM2002two}
A.~Lodi, S.~Martello, and M.~Monaci.
\newblock Two-dimensional packing problems: {A} survey.
\newblock {\em European Journal of Operational Research}, 141(2):241–252,
  2002.

\bibitem{Challenge24_2}
C.~Luo, Z.~Su, and Z.~Lü.
\newblock A general heuristic approach for maximum polygon packing.
\newblock In {\em Symposium on Computational Geometry (SoCG)}, volume 293 of
  {\em LIPIcs}, pages 84:1--84:9, 2024.

\bibitem{merino2020two}
A.~Merino and A.~Wiese.
\newblock On the two-dimensional knapsack problem for convex polygons.
\newblock {\em ACM Transactions on Algorithms}, 2020.

\bibitem{milenkovic1997multiple}
V.~Milenkovic.
\newblock Multiple translational containment part ii: Exact algorithms.
\newblock {\em Algorithmica}, 19(1):183--218, 1997.

\bibitem{milenkovic1996translational}
V.~J. Milenkovic.
\newblock Translational polygon containment and minimal enclosure using linear
  programming based restriction.
\newblock In {\em Symposium on Theory of Computing (STOC)}, pages 109--118,
  1996.

\bibitem{milenkovic1998rotational}
V.~J. Milenkovic.
\newblock Rotational polygon containment and minimum enclosure.
\newblock In {\em Symposium on Computational Geometry (SoCG)}, pages 1--8,
  1998.

\bibitem{milenkovic1999rotational}
V.~J. Milenkovic.
\newblock Rotational polygon containment and minimum enclosure using only
  robust 2d constructions.
\newblock {\em Computational Geometry}, 13(1):3--19, 1999.

\bibitem{milenkovic1999translational}
V.~J. Milenkovic and K.~Daniels.
\newblock Translational polygon containment and minimal enclosure using
  mathematical programming.
\newblock {\em International Transactions in Operational Research},
  6(5):525--554, 1999.

\bibitem{SoCG2020_2}
L.~Moalic, D.~Schmitt, J.~Lepagnot, and J.~Kritter.
\newblock Computing low-cost convex partitions for planar point sets based on a
  memetic approach.
\newblock In {\em Symposium on Computational Geometry (SoCG)}, volume 164 of
  {\em LIPIcs}, pages 84:1--84:9, 2020.

\bibitem{MM1967some}
J.~W. Moon and L.~Moser.
\newblock Some packing and covering theorems.
\newblock In {\em Colloquium Mathematicae}, volume~17, pages 103--110, 1967.

\bibitem{area-campinas}
N.~Ramos, R.~C. de~Jesus, P.~de~Rezende, C.~de~Souza, and F.~L. Usberti.
\newblock Heuristics for area optimal polygonizations.
\newblock {\em Journal of Experimental Algorithmics}, 27:2.1:1--2.1:25, 2022.

\bibitem{SoCG2022_4}
A.~Schidler.
\newblock Sat-based local search for plane subgraph partitions.
\newblock In {\em Symposium on Computational Geometry (SoCG)}, volume 224 of
  {\em LIPIcs}, pages 74:1--74:8, 2022.

\bibitem{Challenge2022_2}
A.~Schidler and S.~Szeider.
\newblock Sat-boosted tabu search for coloring massive graphs.
\newblock {\em Journal of Experimental Algorithms}, 28:1.5:1--1.5:19, 2023.

\bibitem{SoCG2022_2}
J.~Spalding{-}Jamieson, B.~Zhang, and D.~W. Zheng.
\newblock Conflict-based local search for minimum partition into plane
  subgraphs.
\newblock In {\em Symposium on Computational Geometry (SoCG)}, volume 224 of
  {\em LIPIcs}, pages 72:1--72:6, 2022.

\bibitem{sweeney1992cutting}
P.~E. Sweeney and E.~R. Paternoster.
\newblock Cutting and packing problems: a categorized, application-orientated
  research bibliography.
\newblock {\em Journal of the Operational Research Society}, 43(7):691--706,
  1992.

\bibitem{cgal}
{The CGAL Project}.
\newblock {\em {CGAL} User and Reference Manual}.
\newblock {CGAL Editorial Board}, {5.5.2} edition, 2023.

\bibitem{toth20172}
G.~F. T{\'o}th.
\newblock Packing and covering.
\newblock In {\em Handbook of Discrete and Computational Geometry, Third
  Edition}, pages 27--66. Chapman and Hall/CRC, 2017.

\bibitem{SoCG2021_3}
H.~Yang and A.~Vigneron.
\newblock A simulated annealing approach to coordinated motion planning.
\newblock In {\em Symposium on Computational Geometry (SoCG)}, volume 189 of
  {\em LIPIcs}, pages 65:1--65:9, 2021.

\bibitem{Challenge2021_3}
H.~Yang and A.~Vigneron.
\newblock Coordinated path planning through local search and simulated
  annealing.
\newblock {\em Journal of Experimental Algorithms}, 27:3.3:1--3.3:14, 2022.

\bibitem{SoCG2020_1}
D.~W. Zheng, J.~Spalding{-}Jamieson, and B.~Zhang.
\newblock Computing low-cost convex partitions for planar point sets with
  randomized local search and constraint programming.
\newblock In {\em Symposium on Computational Geometry (SoCG)}, volume 164 of
  {\em LIPIcs}, pages 83:1--83:7, 2020.

\end{thebibliography}
\end{document}